\documentstyle[preprint,tighten,aps,epsf]{revtex}
\def\lsim{\raise0.3ex\hbox{$\;<$\kern-0.75em\raise-1.1ex\hbox{$\sim\;$}}}
\def\gsim{\raise0.3ex\hbox{$\;>$\kern-0.75em\raise-1.1ex\hbox{$\sim\;$}}}
\begin{document}
\preprint{SISSA/107/99/EP ;  IC/99/124}
\draft
\title{CP conserving constraints on Supersymmetric CP violation in the MSSM}
\author{D. A. Demir$^{a}$, A. Masiero$^{b}$, O. Vives$^{b}$}
\address{$^{a}$ The Abdus Salam International Center for Theoretical Physics,
I-34100, Trieste, Italy }
\address{$^{b}$ SISSA -- ISAS, Via Beirut 4, I-34013, Trieste, Italy and \\ 
INFN, Sezione di Trieste, Trieste, Italy.}
\maketitle

\begin{abstract}
We address the following question. Take the Constrained Minimal 
Supersymmetric Standard Model (CMSSM) with the two CP violating SUSY phases
different from zero, and neglect the bound coming from the electric dipole 
moment (EDM) of the neutron: is it possible to fully account for CP violation 
in the kaon and B systems using only the SUSY contributions with a vanishing 
CKM phase? We show that the $\mbox{BR}(B\rightarrow X_{s} \gamma)$ constraint,
though CP conserving, forces a negative answer to the above question.
This implies that even in the regions of the CMSSM where a cancellation of 
different contributions to the EDM allows for large SUSY phases, it is not 
possible to exploit the SUSY phases to fully account for observable CP 
violation. Hence to have sizeable SUSY contributions to CP violation, one needs
new flavor structures in the sfermion mass matrices beyond the usual CKM 
matrix.
\end{abstract}
\pacs{13.25.Es, 13.25.Hw, 11.30.Er, 12.60.Jv}

\section{Introduction}

Since the initial work of Kobayashi and Maskawa, the Standard Model (SM) of 
electroweak interactions is known to be able to accommodate the experimentally 
observed CP--violation through a unique phase, $\delta_{CKM}$, in the
Cabibbo--Kobayashi--Maskawa mixing matrix (CKM). However, the available 
experimental information, namely $\varepsilon_K$ and 
$\varepsilon^\prime/\varepsilon$, is not enough to establish this phase as 
the only source of CP--violation.

Most of the extensions of the SM include new observable phases
that may significantly modify the pattern of CP violation.
Supersymmetry is, without a doubt, one of the most popular extensions of the 
SM. 
Indeed, in the minimal supersymmetric extension of the SM (MSSM), there 
are additional phases which can cause deviations from the predictions of the 
SM. After all possible rephasings of the parameters and fields, there remain 
at least two new physical phases in the  MSSM Lagrangian. These phases can be 
chosen to be the phases of the Higgsino Dirac mass parameter 
($\phi_{\mu}=\mbox{Arg}[\mu]$) and  the trilinear sfermion coupling 
to the Higgs, ($\phi_{A_{0}}=\mbox{Arg}[A_{0}]$) 
\cite{2phases}. In fact, in the so--called Constrained Minimal Supersymmetric 
Standard Model (CMSSM), with strict universality at the Grand Unification 
scale, these are the only new phases present.

It was soon realized that for most of the CMSSM parameter space, the 
experimental bounds on the electric dipole moments of the electron and 
neutron constrained $\phi_{A_0,\mu}$ to be at most ${\cal{O}}(10^{-2})$.
Consequently these new supersymmetric phases have been taken to vanish 
exactly in most studies of CMSSM.
  
However, in the last few years, the possibility of having non-zero SUSY phases
has again attracted a great deal of attention. Several new mechanisms have 
been proposed to suppress EDMs below the experimental bounds while allowing 
SUSY phases ${\cal{O}}(1)$. Methods of suppressing the EDMs 
consist of cancellation of various SUSY contributions among themselves 
\cite{cancel}, non-universality of the soft breaking parameters at the 
unification scale \cite{non-u} and approximately degenerate heavy sfermions 
for the first two generations \cite{heavy}. 
In a recent work \cite{fully}, we showed that, in a model with heavy 
sfermions of the first two generations and in the large $\tan \beta$ regime,
$\varepsilon_K$ and $\varepsilon_B$ could receive very sizeable contributions 
from these new SUSY phases. Similar studies \cite{korea} including a larger 
set of experimental constraints have reported the impossibility of such 
large supersymmetric contributions \footnote{In this paper we restrict our
discussions to the CMSSM. If one relaxes some of the constraints of this model,
for instance by allowing for large gluino mediated CP violation with 
non--universal soft SUSY breaking terms, then it might still be possible 
to have fully supersymmetric $\varepsilon$ and 
$\varepsilon^{\prime}/\varepsilon$ \cite{new-korean}.}.

In this work, we are going to complete our previous analysis 
with the inclusion of all the relevant constraints in a CMSSM scenario.
In doing so we adopt a different perspective. We will assume from the very 
beginning that both supersymmetric phases are ${\cal{O}}(1)$, ignoring for the 
moment EDM bounds \footnote{ EDM cancellations may be obtained through 
non--trivial relative phases in the gaugino mass parameters (see for instance 
the third paper in Ref. \cite{cancel}). However, for the discussion of the 
present paper, no explicit mechanism for such a cancellation is needed}. 
In these conditions, and taking into account other 
CP--conserving constraints, we will analyze the effects on the low energy 
CP--violation observables, especially $\varepsilon_K$ and $\varepsilon_B$.
It should be noticed that the model used in \cite{fully} can be easily 
obtained as a limit of the CMSSM by decoupling the first two generations of
squarks and neglecting the intergeneration mixing in the sfermion mass 
matrices. Hence, our results in the more general CMSSM will include this model
as a limiting case.

In the next section we study the new sources of flavor mixing 
present at the electroweak scale in any supersymmetric model.  
In section 3 we are going to analyze neutral meson mixing, i.e. 
$K^0$--$\bar{K}^0$ and $B^0$--$\bar{B}^0$ mixings, with large supersymmetric 
phases. Section 4 will be devoted to the study of the branching ratio of 
the decay $b \rightarrow s \gamma$. In section 5 we will show the
impact of the measured $b \rightarrow s \gamma$ branching ratio on the 
supersymmetric contributions to $\varepsilon_K$ and $\varepsilon_B$.
Section 6 will present our conclusions and in Appendices A and B 
we collect, respectively, the formulas for the integration of relevant RGE's and
the different loop functions appearing in the text.

\section{Flavor mixing in the CMSSM}
The issue of flavor changing neutral current (FCNC) interactions in the CMSSM 
has been widely investigated in 
the literature. For the completeness of the discussion, we briefly recall 
those properties which will be relevant for our analysis. 
 
The  CMSSM is the simplest Supersymmetric structure we can build from the SM 
particle content. This model is completely defined once we specify the 
soft-supersymmetry breaking terms. These are taken to be
strictly universal at some ultra-high energy scale, which we take to be the 
Grand Unification scale ($M_{GUT}$). 

\begin{eqnarray}
\label{soft}
& (m_Q^2)_{i j} = (m_U^2)_{i j} = (m_D^2)_{i j} = (m_L^2)_{i j}
 = (m_E^2)_{i j} = m_0^2\  \delta_{i j}, & \nonumber \\
& m_{H_1}^2 = m_{H_2}^2 = m_0^2, &\\
& m_{\tilde{g}}\ =\ m_{\tilde{W}}\ =\ m_{\tilde{B}}\ =\ m_{1/2}, &  
\nonumber \\
& (A_U)_{i j}= A_0 e^{i \phi_A} (Y_U)_{i j},\ \ \ (A_D)_{i j}= A_0 e^{i \phi_A}
(Y_D)_{i j},\ \ \ 
(A_E)_{i j}= A_0e^{i \phi_A} (Y_E)_{i j}. & \nonumber 
\end{eqnarray}

That is, there is a common mass for all the scalars, $m_0^2$, a single gaugino 
mass, $m_{1/2}$, and all the trilinear soft-breaking terms are directly 
proportional to the corresponding Yukawa couplings in the superpotential 
with a proportionality constant $A_0 e^{i \phi_A} $.

Now, with the use of the Renormalization Group Equations (RGE) of the MSSM,
as explained in Appendix A, we can obtain the whole Supersymmetric spectrum 
at the electroweak scale. All the Supersymmetric masses and mixings are then
a function of $m_0^2$, $m_{1/2}$, $A_0$, $\phi_A$, $\phi_\mu$ and 
$\tan \beta$. We require radiative symmetry breaking to fix $|\mu|$ and 
$|B \mu|$ \cite{rge,bertolini} with tree--level Higgs potential
\footnote{The RGE's of the MSSM have received a vast amount of attention in 
the literature. However, in most of the previous analyses the SUSY phases 
$\phi_{A}$ and $\phi_{\mu}$ are switched off. For this reason we prefer 
to give the relevant RGE's with non--vanishing SUSY phases in Appendix A.}.

It is important to notice that, even in a model with universal soft breaking
terms at some high energy scale as the CMSSM, some off--diagonality in
the squark mass matrices appears at the electroweak scale. Working on the 
basis where the squarks are rotated parallel to the quarks, the so--called 
Super CKM basis (SCKM), the squark mass matrix is not flavor diagonal at 
$M_W$. This is due to the fact that at $M_{GUT}$ there exist two non-trivial 
flavor structures, namely the two Yukawa matrices for the up and down quarks, 
which are not simultaneously diagonalizable. This implies that 
through RGE evolution some flavor mixing leaks into the sfermion mass matrices.
In a general Supersymmetric model, the presence of new flavor structures
in the soft breaking terms would generate large flavor mixing in the sfermion 
mass matrices. However, in the CMSSM, the two Yukawa matrices are the only 
source of flavor change. As always in the SCKM basis, any off-diagonal entry 
in the sfermion mass matrices at $M_W$ will be necessarily proportional to a 
product of Yukawa couplings. The RGE's for the soft breaking terms
are sets of linear equations, and thus, to match the correct quirality of the 
coupling, Yukawa couplings or tri-linear soft terms must enter the RGE in 
pairs, as we can see in Eqs. (A.1-A.3) in Appendix A.

In fact, in the up (down) squark mass matrix the up (down) Yukawas will 
also be diagonalized and so will mainly contribute to diagonal entries 
while off--diagonal entries will be due to the down (up) Yukawa matrix.
This means, for instance, that in this model the off-diagonality in the 
$M^{(d)\ 2}_{LL}$ matrix will roughly be $c \cdot Y_u Y_u^\dagger$. With $c$ a
proportionality factor that typically is, 
\begin{eqnarray}
c \simeq \frac{1}{(4 \pi)^2} \log \left(\frac{M_{Gut}}{M_W}\right) \simeq 0.20
\end{eqnarray}
as expected from the loop factor and the running from $M_{GUT}$ to $M_W$.
Nevertheless, we have to keep in mind that this is simply a typical estimate 
and the final value of $c$ can suffer a sizeable variation depending on 
many other factors not present in this simple estimate.

On the other hand, this has clear implications on the $\tan \beta$ dependence
of these off--diagonal entries of the sfermion mass matrices.
In the basis where the down Yukawa matrix is diagonal, we can write the up 
and down Yukawas as,
\begin{eqnarray}
\label{iki}
Y_{U}(M_{Z})=\frac{g}{\sqrt{2} M_{W} \sin\beta} V^{\dagger}_{CKM}\, M_u\,,
\ \ \ \ \ 
Y_{D}(M_{Z})=\frac{g}{\sqrt{2} M_{W}\cos\beta}\, M_d
\end{eqnarray}
with $V_{CKM}$ the Cabibbo-Kobayashi-Maskawa mixing matrix and $M_{u,d}$ the 
diagonalized mass matrices for the quarks. We can see in this 
equation that, for $\tan \beta \gsim 1$, the up Yukawa matrix will maintain 
similar values when going to large $\tan \beta$. Hence, the off--diagonal 
entries in the down squarks mass matrix will be roughly stable with 
$\tan \beta$.  
In the up squark mass matrix we have the opposite situation and the 
$\tan \beta$ dependence is very strong. In 
this case the off--diagonal entries depend on the down Yukawa 
matrix that grows linearly with $\tan \beta$ for large $\tan \beta$. 
This means that we can expect the flavor change in the up and down squark 
mass matrix to be similar when $\tan \beta \simeq m_t/m_b \simeq 40$.
While for $\tan \beta \simeq 2$ the flavor change in the up mass matrix
will be approximately $(\tan \beta = 40)^2/(\tan \beta = 2)^2 = 400$ times
smaller (see Appendix A for details).  
These points also apply to the left--right sub--matrices where again 
flavor changing entries will be due to the opposite isospin Yukawa matrix.
In fact, this left--right sfermion mixing only appears after electroweak 
symmetry breaking. The expression for these matrices in the SCKM basis is,
\begin{eqnarray}
\label{LR}
{M_{LR}^{(u)}}^{2}&=& \left(\frac{v_2}{\sqrt{2}} V_{CKM} A_{U}^*(M_{Z})-
|\mu (M_{Z})|e^{i\phi_{\mu}}\cot\beta\, M_{u}\right)\\
{M_{LR}^{(d)}}^{2}&=&\frac{v_1}{\sqrt{2}} A_{D}^*(M_{Z})-
|\mu (M_{Z})|e^{i\phi_{\mu}}\tan\beta\, M_{d}
\end{eqnarray}
Then, these left--right mixings will have an additional suppression
proportional to the mass of the corresponding right--handed quark
( remember that $A_U v_1 \approx A_0 M_U$). This is 
always true for all the generation changing entries that are produced by
the $A$ matrices. However, in the down mass matrix, this suppression can be 
partially compensated by a large value of $\tan \beta$ in the diagonal terms 
proportional to $\mu$.  
These are all well--known facts in the different studies of FCNC processes
in the framework of the CMSSM \cite{bertolini,cho} and imply that flavor 
mixing is still dominantly given by the usual CKM mixing matrix in W-boson, 
charged Higgs and chargino vertices. 

In this work, we are especially interested in CP violating observables.  
Then we must also consider the presence of observable phases in the sfermion 
mass matrices. In the following we will take the CKM matrix exactly real
to isolate pure effects of the new supersymmetric phases \cite{new-branco}.
The sfermion mass matrices contain several physical phases that give rise to 
CP violation phenomena. In particular, before RGE evolution, these phases
($\phi_A$, $\phi_\mu$) are confined to the left-right part of
the sfermion mass matrix while both the left--left, $m^2_{Q}$, and 
right--right, $m^2_{U,D}$, matrices are real and diagonal. 
However this is not true anymore at $M_W$: $\phi_A$ leaks into the 
off-diagonal elements of these hermitian matrices through RGE evolution.
From the explicit RGE in the MSSM, Eq.(\ref{mqrge}), it is clear that 
this phase only enters the $(m_Q^2)_{i j}$ evolution through the combinations 
$(A_U A_U^\dagger)_{i j}$ or $(A_D A_D^\dagger)_{i j}$. At $M_{GUT}$ these 
matrices have a common phase, and so the combination $(A A^\dagger)$ is exactly
real. So, to the extent that the $A$ matrices keep a 
uniform phase during RGE evolution, no phase will leak into the $m_{Q}^2$ 
matrices. However, we can easily see from Eqs.(A.2) and (A.3) that this is 
not the case, and different elements of the $A$ matrices are renormalized 
differently. In this equation, we can see that only the 
terms involving two Yukawa and one $A$ matrix can produce a mismatch in the 
phases. Moreover, these terms will only be important when there are no small 
Yukawas involved. Then, we can expect a mismatch only on the off--diagonal 
elements involving the third generation. Keeping this in mind, the general 
form of the $m_{Q}^2$ matrix at $M_W$ in terms of the initial conditions is,
\begin{eqnarray}
m_{Q}^{2}(M_{W})=\eta^{(m)}_{Q} m_{0}^{2}+\eta^{(A)}_{Q} A_{0}^{2}+
\eta^{(g)}_{Q} m_{1/2}^{2}+ \Big(\eta^{(g A)}_{Q} e^{i\phi_{A}} + 
\eta^{(g A)\,T}_{Q} e^{- i\phi_{A}}\Big) A_{0} m_{1/2} 
\end{eqnarray}
where the coefficients $\eta$ are $3\times 3$ matrices with real numerical 
entries. In this expression we can see that the presence of imaginary parts
will be linked to the non-symmetric part of the $\eta^{(g A)}_{Q}$ matrices.
As is clear from the mass matrices in Appendix A (Eqs. (A.5-A.7) and 
(A.12-A.14)), these non--symmetric parts of $m_Q^2$ are always more that 
three orders of magnitude smaller than the corresponding symmetric parts. 
This means that, in the SCKM basis, the imaginary parts of any mass insertion
are present only in one part per $2-3 \times 10^3$, and are always associated 
with $(3,i)$ MI, Eqs. (A.8-A.11) and (A.15-A.18). A very similar situation 
was also found by Bertolini and Vissani, in the CMSSM with vanishing susy 
phases for the leakage of $\delta_{CKM}$ \cite{vissani,non-u}. So, we 
conclude that in the processes we will consider, we can take both 
${M^{(u)}}^{2}_{LL}$ and ${M^{(d)}}^{2}_{LL}$ as real to a very good 
approximation.   

In the following we will analyze the new effects of this model on 
indirect CP violation in $K$ and $B$ systems. In doing so, we will use
both the exact vertex mixing method and the Mass Insertion (MI) approximation
\cite{MI}. Notice that the MI approximation is extremely good in the case of 
the CMSSM where all the off-diagonal entries are sufficiently small. The 
size of these off--diagonal entries directly gives in the MI approximation  
the amount of flavor changing induced by the sfermion mass matrices. A possible
exception may arise in the stop and sbottom sectors that, in any case, could 
be diagonalized to ensure the validity of the MI approximation \cite{MI-2}.
As we will see in the next sections, this is frequently useful to understand 
the exact results obtained in the vertex mixing method.  

\section{Indirect CP violation in the CMSSM}  

In the SM neutral meson mixing arises at one loop through the well--known
$W$--box. However, in the CMSSM, there are new contributions to $\Delta F=2$
processes coming from  boxes mediated by supersymmetric particles. These are
charged Higgs boxes ($H^{\pm}$), chargino boxes ($\chi^{\pm}$) and 
gluino-neutralino boxes ($\tilde{g}$, $\chi^{0}$). The amount of the indirect
CP violation in the neutral meson ${\cal{M}}$ system is measured by 
the well--known $\varepsilon_{{\cal{M}}}$ parameter:
\begin{eqnarray}
\label{ek}
\varepsilon_{{\cal{M}}}=\frac{1}{\sqrt{2}}\frac{{\cal{I}}m\langle {\cal{M}}^0|
{\cal H}_{eff}^{\Delta F=2} |\bar{{\cal{M}}}^0 \rangle}{\Delta M_{\cal{M}}}
\end{eqnarray}
where $\Delta M_{\cal{M}}$ is the ${\cal{M}}$--${\cal{\overline{M}}}$ mass 
splitting.
$\varepsilon_{{\cal{M}}}$ depends on the matrix elements of the $\Delta F=2$
Hamiltonian, ${\cal{H}}_{eff}^{\Delta F=2}$, which can be decomposed as
\begin{eqnarray}
\label{DF=2}
{\cal{H}}_{eff}^{\Delta F=2}=-\frac{G_{F}^{2} M_{W}^{2}}{(2 \pi)^{2}}
(V_{td}^{*} V_{tq})^{2}( C_{1}(\mu) Q_{1}(\mu)
+C_{2}(\mu) Q_{2}(\mu) +C_3(\mu) Q_3(\mu))
\end{eqnarray}
where the relevant four--fermion operators are given by 
\begin{eqnarray}
Q_{1}&=&\bar{d}^{\alpha}_{L}\gamma^{\mu}q^{\alpha}_{L}\cdot 
\bar{d}^{\beta}_{L}\gamma_{\mu}q^{\beta}_{L},\nonumber\\ 
Q_{2}&=&\bar{d}^{\alpha}_{L}q^{\alpha}_{R}\cdot \bar{d}^{\beta}_{L}
q^{\beta}_{R},\nonumber\\ 
Q_{3}&=&\bar{d}^{\alpha}_{L}q^{\beta}_{R}\cdot \bar{d}^{\beta}_{L}
q^{\alpha}_{R}
\end{eqnarray}
with $q=s , b$ for the $K$ and $B$--systems respectively and $\alpha, 
\beta$ as color indices. In the CMSSM, these are the only three operators
present in the limit of vanishing $m_d$.

At this point, we are going to divide our discussion into two parts. We 
analyze separately the effective operator $Q_1$ that preserves chirality 
along the fermionic line, and the operators $Q_2$ and $Q_3$ that change 
chirality along the fermionic line. As we will see below the flavor mixing 
in the sfermion mass matrix and the experimental constraints on both kinds of
operators are very different.

\subsection{Chirality conserving transitions} 

In Eq.(\ref{DF=2}), $Q_1$ is the only operator present that does not involve 
a chirality change in the fermionic line. With respect to the associated
sfermion, no chirality change in the sfermion propagator 
will be needed, and so, the suppression associated with left--right sfermion
mixing can be avoided. In general, ${C_{1}}(\mu_{0})$ can be decomposed as 
follows
\begin{eqnarray}
{C_{1}}(\mu_{0})&=&{C_{1}^{W}}(M_W)+{C_{1}^{H}}(M_W)+
C_{1}^{\tilde{g},\chi^0}(M_W)+ {C_{1}^{\chi}}(M_W)\,.
\end{eqnarray} 

The usual SM $W$--box, where all the couplings are purely left--handed can 
only contribute to this effective operator. However, with $\delta_{CKM}=0$,
$C_1^W$ does not contain any complex phase and hence cannot contribute to the 
imaginary part in $\varepsilon_{\cal{M}}$.
In any case it will always be, in the CMSSM, the dominant contribution to
$\Delta M_{\cal{M}}$. Similarly, the charged Higgs contribution, 
$C_1^H$ depends on the same combination of CKM elements with no other
CP--violating phase \cite{bertolini}. So it will not contribute to our CP 
violating observable.

Gluino and neutralino contributions to $C_1^{\tilde{g},\chi^0}$ are 
specifically supersymmetric. They involve the superpartners of 
quarks and gauge bosons. Here, the source of 
flavor mixing is not directly the usual CKM matrix. It is the presence of
off--diagonal elements in the sfermion mass matrices, as discussed in the 
previous section. From the point of view of CP violation, we will always need
a complex Wilson coefficient. In the SCKM basis all gluino vertices are 
flavor diagonal and real. This means that in the MI approximation,
we need a complex mass insertion in one of the sfermion lines. As explained 
in the previous section, these MI are proportional to Yukawa couplings
and real up to 1 part in $2 \times 10^3$. The complete 
expressions for the gluino contributions to $\Delta F=2$ processes in the MI 
approximation can be found in \cite{gabbiani}. The bounds obtained there
for the real and imaginary parts of the mass insertions required to saturate 
$\Delta M_K$ and $\varepsilon_K$ are,
\begin{eqnarray}
\label{bounds}
\sqrt{|Re (\delta^d_{12})^2_{LL}|} < 4 \cdot 10^{-2} \\
|(\delta^d_{1 2})_{L L}| \sin (2\phi_{L L}) < 3 \cdot 10^{-3} \nonumber \\
(\delta^d_{i j})_{A B} = \frac{(M_{AB}^2)_{i j}}{\tilde{M}} \nonumber
\end{eqnarray}   
where $\tilde{M}$ is an average squark mass.

In the CMSSM, as we can see in Appendix A, these mass insertions are 
much smaller. In particular, the fact that the bound on $\Delta M_K$ ,the 
real part of the MI, is satisfied implies that the imaginary parts are at 
least two orders of magnitude below the required value to saturate 
$\varepsilon_K$ . Hence, no sizeable contributions to $\varepsilon_K$ from 
gluino boxes are possible. The situation in 
$B^0$--$\bar{B}^0$ mixing is completely analogous: assuming that the minimum
phase from the mixing observable in the B--factories is around $0.1$ radians,
we would need an imaginary contribution not more than one order of magnitude 
below the real one. With the arguments
given above, this is clearly out of reach for gluino boxes in the
CMSSM. Neutralino contributions are generally smaller than gluino due to
smaller couplings with the same source of flavor mixing. In fact, although  
neutralino vertices in the SCKM basis also involve the complex neutralino 
mixings, any imaginary part on this operator will only be due to a complex 
mass insertion. This can be seen in the explicit expressions in 
\cite{bertolini} where all neutralino mixings in this operator appear in pairs
with its complex conjugate counterpart. 

Finally, the charginos also contribute to ${C_{1}}(M_W)^\chi$. In this case,
flavor mixing comes explicitly from the CKM mixing matrix, although
off--diagonality in the sfermion mass matrix introduces a small additional
source of flavor mixing.  

\begin{eqnarray}
\label{chWC}
C_1^\chi (M_W) = \sum_{i,j=1}^{2} \sum_{k, l=1}^{6} 
\sum_{\alpha \gamma \alpha^\prime \gamma^\prime} \frac{V_{\alpha^\prime d}^{*} 
V_{\alpha q}V_{\gamma^\prime d}^{*} V_{\gamma q}}{(V_{td}^{*} V_{tq})^2}
[G^{(\alpha,k)i} G^{(\alpha^\prime,k)j*} G^{(\gamma^\prime,l)i*} 
G^{(\gamma,l)j}\  Y_1(z_{k}, z_{l}, s_i, s_j)] 
\end{eqnarray}
where $z_k = M_{\tilde{u}_k}^2/M_W^2$, $s_i =M_{\tilde{\chi}_i}^2/M_W^2$, and 
$V_{\alpha q} G^{(\alpha,k)i}$ represent the coupling of chargino, $i$, and 
squark $k$ to the left--handed down quark, $q$. Finally $\alpha$ is an 
intermediate up--quark index associated with the factorization of the CKM 
mixing matrix. The expression for this coupling is then,
\begin{eqnarray}
G^{(\alpha, k) i}&=& \left( \Gamma_{U L}^{\alpha k} C_{R 1 i}^{*} - 
\frac{m_{\alpha}}{\sqrt{2} M_W \sin \beta} \Gamma_{U R}^{\alpha k} 
C_{R 2 i}^{*}\right)
\end{eqnarray}
where $\Gamma_{U L}$ and $\Gamma_{U R}$ are $6\times 3$ matrices such that 
the $6\times 6$ unitary matrix $\Gamma_{U}\equiv \{\Gamma_{U L}\Gamma_{U R}\}$ 
diagonalizes the up--squark mass matrix, $\Gamma_{U} M_{U}^{2}
\Gamma_{U}^{\dagger}=diag(M_{\tilde{u}_{1}}^{2}, ... ,M_{\tilde{u}_{6}}^{2})$.
$C_R$ is one of the matrices that diagonalize the chargino mass matrix through
a bi--unitary transformation $C_{R}^{\dagger} M^-_{\chi} 
C_{L}=diag(M_{\chi^{\pm}_{1}}, M_{\chi^{\pm}_{2}})$, with,
\begin{eqnarray}
\label{chi-}
M^{-}_{\chi}&=&\left (\begin{array}{c c}
\tilde{m}_W & M_W \cos \beta\\
M_W \sin \beta & |\mu| e^{i \phi_\mu} \end{array}\right)
\end{eqnarray}
From these equations it is clear that $G^{(\alpha, k) i}$ will in general be 
complex, as both $\phi_\mu$ and $\phi_A$ are present in the different mixing 
matrices. The loop function $Y_1(a, b, c, d)$ is given in Eq. (B.1) of 
Appendix B.

\begin{figure}
\begin{center}
\epsfxsize = 15cm
\epsffile{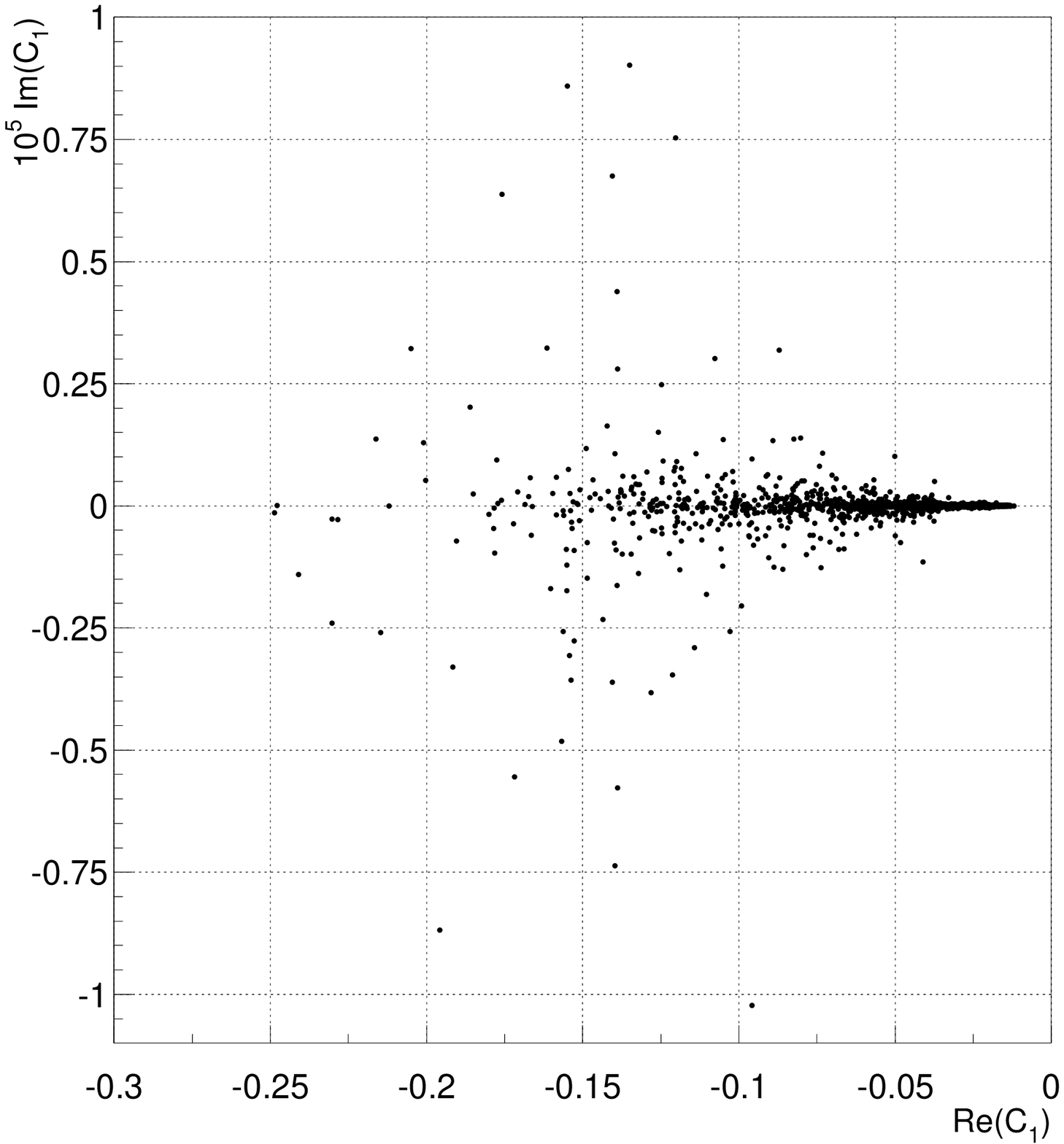}
\leavevmode
\end{center}
\caption{Imaginary and Real parts of the Wilson coefficient $C_1^\chi$ in
Kaon mixing.}
\label{imK40}
\end{figure}

\begin{figure}
\begin{center}
\epsfxsize = 15cm
\epsffile{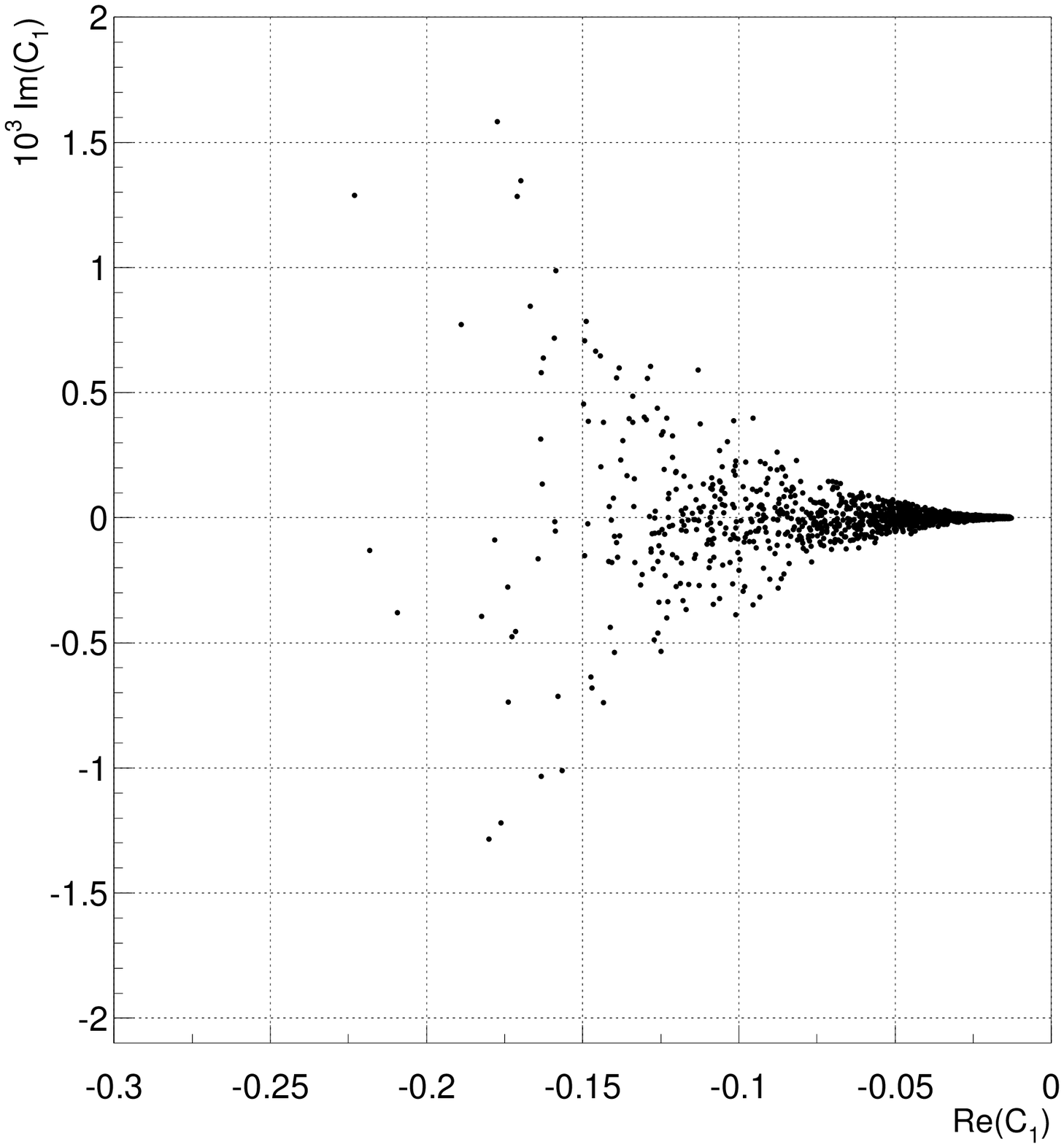}
\leavevmode
\end{center}
\caption{Imaginary and Real parts of the Wilson coefficient $C_1^\chi$ in
B mixing.}
\label{imB40}
\end{figure}

The main part of $C_1^\chi$ in Eq(\ref{chWC}) will be given by pure CKM 
flavor mixing, neglecting the additional flavor mixing in the squark
mass matrix \cite{cho,branco}. In this case, $\alpha= \alpha^\prime$ and
$\gamma=\gamma^\prime$, we have,
\begin{eqnarray}
\label{chWC1real}
C_1^{(0)\chi} (M_W) = \sum_{i,j=1}^{2} \sum_{k, l=1}^{6} 
\sum_{\alpha \gamma} \frac{V_{\alpha d}^{*} 
V_{\alpha q}V_{\gamma d}^{*} V_{\gamma q}}{(V_{td}^{*} V_{tq})^2}
[G^{(\alpha,k)i} G^{(\alpha,k)j*} G^{(\gamma,l)i*} 
G^{(\gamma,l)j}\  Y_1(z_{k}, z_{l}, s_i, s_j)] 
\end{eqnarray}  
But, taking into account that $ Y_1(a,b,c,d)$ is symmetric 
under the exchange of any pair of arguments we have,
\begin{eqnarray}
\label{real}
&G^{(\alpha,k)i} G^{(\alpha,k)j*} G^{(\gamma,l)i*} 
G^{(\gamma,l)j}\  Y_1(z_{k}, z_{l}, s_i, s_j)=&\\
& \frac{1}{2}\left(G^{(\alpha,k)i} 
G^{(\alpha,k)j*} G^{(\gamma,l)i*} G^{(\gamma,l)j} + G^{(\alpha,k)i*} 
G^{(\alpha,k)j} G^{(\gamma,l)i} G^{(\gamma,l)j*}\right) 
Y_1(z_{k}, z_{l}, s_i, s_j)& \nonumber
\end{eqnarray}
and so $C_1^{(0)\chi}$, is exactly real \cite{fully}. 
This is not exactly true in the CMSSM, where there is additional flavor change 
in the sfermion mass matrices. Here, some imaginary parts appear in the 
$C_1^\chi$ in Eq(\ref{chWC}). Being associated to the size of 
intergenerational sfermion mixings, these imaginary parts will be maximal
for large $\tan \beta$. In Fig(\ref{imK40}) we show in a scatter plot the 
size of imaginary and real parts of $C_1^\chi$ in the K system for a fixed 
value of $\tan \beta=40$. The region of susy parameters explored in this 
and all of the following scatter plots is $50\ GeV \leq m_0, m_{1/2}, A_0 \leq
500\ GeV$ and $0 \leq \phi_A, \phi_\mu \leq 2 \pi$. With these initial 
conditions we impose that all squarks are heavier than $100\ GeV$ with the 
exception of the stops that, as the charginos, are only required to be 
above $80\ GeV$. Furthermore we impose the constraint from the 
$b \rightarrow s \gamma$ decay. Notice that, as 
we will see later, this is a conservative attitude, in the sense that other 
constraints that we do not impose could only make our conclusions stronger.  
Under these conditions, we can see here that, in the CMSSM, this Wilson 
coefficient is always real up to a part in $10^5$.
Fig(\ref{imB40}) is the equivalent plot for the case of $B^0$--$\bar{B}^0$ 
mixing. Here, imaginary parts are relatively larger but, in any case,
out of reach for the foreseen B--factories.

Taking this into account, from the point of view of experimental interest,
we will always neglect imaginary parts in the Wilson Coefficient $C_1$ 
within the CMSSM. Notice that this would not apply in a general model
with non-universality at the GUT scale \cite{gabbiani}, and each particular 
model should be considered separately.

\subsection{Chirality changing transitions} 

From the point of view of flavor change and CP violation, operators $Q_2$ and 
$Q_3$ are different from $Q_1$. These two operators always involve a change 
in the chirality of the external quarks, and consequently also a change of the 
chirality of the associated squarks or gauginos. In particular, this implies
the direct involvement of the supersymmetric phases. On the other hand, these 
operators are suppressed by the presence of down quark Yukawa couplings, and 
so can only be relevant in the region of large $\tan \beta$ \cite{fully}. 
We can write the different contributions to $C_2$ and $C_3$ as,
\begin{eqnarray}
{C_{2}}(M_W)&=&{C_{2}^{H}}(M_W) + C_{2}^{\tilde{g}}(M_W) \\
{C_{3}}(M_W)&=& C_{3}^{\tilde{g},\chi^0}(M_W)+ {C_{3}^{\chi}}(M_W) \nonumber
\end{eqnarray} 
In first place, charged Higgs contributes only to $C_2$, but parallel to the 
discussion for $C_1^{W,H}$, the absence of phases prevents it from 
contributing to $\varepsilon_{\cal M}$.
 
Gluino and neutralino boxes contribute both to $Q_2$ and $Q_3$. However 
flavor change will be given in this case by an off--diagonal left--right mass 
insertion. In the CMSSM these MI are always proportional to the mass and are
never enhanced by large $\tan \beta$ values (see Eq.(\ref{LR})) of the 
right handed squark. 
This implies that these left--right flavor transitions from gluino will always 
be smaller in the CMSSM than the corresponding chargino contributions, where 
flavor change is directly given by the CKM matrix. In fact, this is already 
well--known for the case of $b \rightarrow s \gamma$ decay \cite{bsg}, 
which is completely equivalent from the point of view of flavor change. 

Hence, the most important contribution, especially for light stop and 
chargino, will be the chargino box. Before the inclusion of QCD effects, it 
contributes solely to the coefficient $C_3$,
  
\begin{eqnarray}
\label{chWCR}
C_3^\chi (M_W) = \sum_{i,j=1}^{2} \sum_{k, l=1}^{6} \sum_{\alpha \gamma
\alpha^\prime \gamma^\prime} 
\frac{V_{\alpha^\prime d}^{*} V_{\alpha q}V_{\gamma^\prime d}^{*} 
V_{\gamma q}}{(V_{td}^{*} V_{tq})^2} \frac{m_q^2}{2 M_W^2 \cos^2 \beta} 
H^{(\alpha,k)i} G^{(\alpha^\prime,k)j*} G^{(\gamma^\prime,l)i*}  
H^{(\gamma,l)j}\nonumber \\
Y_2(z_k, z_l, s_i, s_j) 
\end{eqnarray} 
where $m_q/(\sqrt{2} M_W \cos \beta) \cdot V_{\alpha q} \cdot H^{(\alpha,k)i}$ 
is the coupling of chargino, $i$, and squark, $k$, to the right--handed down 
quark $q$, with,
\begin{equation}
\label{couplingR}
H^{(\alpha,k)i} = C^*_{L 2 i} \Gamma_{U L}^{\alpha k}
\end{equation} 
and $Y_2(a,b,c,d)$ is given in Eq. (B.2).
Unlike the $C_1^\chi$ Wilson coefficient, $C_3^\chi$ is complex even in the 
absence of intergenerational mixing in the sfermion mass matrices \cite{fully}.
In fact, the presence of flavor violating entries in the up--squark mass
matrix hardly modifies the results obtained in their absence 
\cite{cho,branco}. So, in these conditions we have,
\begin{eqnarray}
\label{chWCR0}
C_3^\chi (M_W) = \sum_{i,j=1}^{2} \sum_{k,l =3,6} 
[ F_s(3,k,3,l,i,j)- 2 F_s(3,k,1,1,i,j)+F_s(1,1,1,1,i,j)]\\
F_s(\alpha,k,\gamma,l,i,j)= \frac{m_q^2}{2 M_W^2 \cos^2 \beta} 
H^{(\alpha,k)i}G^{(\alpha,k)j*} 
G^{(\gamma,l)i*} H^{(\gamma,l)j} Y_2(z_{k}, z_{l}, s_i, s_j) 
\nonumber
\end{eqnarray}
where we have used CKM unitarity and degeneracy of the first two generations 
of squarks. Due to the differences between $H$ and $G$ couplings, this 
contribution is always complex in the presence of susy phases. The most
relevant feature of Eqs.(\ref{chWCR}) and (\ref{chWCR0}) is the explicit 
presence of the external quark Yukawa coupling squared, 
$m_q^2/(2 M^2_W \cos^2 \beta)$. This is the reason why this contribution is
usually neglected in the literature \cite{korea,bertolini,branco}.
However, as we showed in \cite{fully}, this contribution could be relevant
in the large $\tan \beta$ regime. For instance, in $B^0$--$\bar{B}^0$ mixing
we have $m_b^2/(2 M^2_W \cos^2 \beta)$ that for $\tan \beta \gsim 25$ is
larger than 1 and so, it is not suppressed at all when compared with the 
$C_1^\chi$ Wilson Coefficient. This means that this contribution can be very 
important in the large $\tan \beta$ regime \cite{fully} and could have 
observable effects in CP violation experiments in the new B--factories.   
However, in our previous work \cite{fully}, we did not include the 
additional constraints coming from $b \rightarrow s \gamma$ decay.
In the next sections we will analyze the relation of $\varepsilon_{\cal M}$
with this decay, and the constraints imposed by its experimental measure. 
       
\section{$b \rightarrow s \gamma$ in the CMSSM}

The decay $b \rightarrow s \gamma$ has already been extensively studied in the 
context of the CMSSM with vanishing susy phases \cite{bsg}. Because the 
branching ratio is a CP conserving observable, the presence of new phases will 
not modify the main features found in \cite{bsg} concerning the relative
importance of the different contributions.
However, in the presence of the new susy phases, these contributions
will have different phases and will be observable through the interference. 
As we will see next, the experimental constraints will also have a large 
impact on the imaginary parts of the decay amplitudes.

This decay is described by the following $\Delta F=1$ effective Hamiltonian 
\begin{eqnarray}
{\cal H}_{eff}^{\Delta F=1}=-\frac{4 G_{F}}{\sqrt{2}} V_{t s}^{*}V_{t b}
\sum_{i=2,7,8} {\cal{C}}_{i} {\cal{Q}}_{i}
\end{eqnarray}
where the relevant operators are given by 
\begin{eqnarray}
{\cal{Q}}_{2}&=&\bar{s}_{L} \gamma_{\mu} c_{L} \bar{c}_{L} \gamma^{\mu} 
b_{L},\\
{\cal{Q}}_{7}&=&\frac{e m_{b}}{16 \pi^{2}}\bar{s}_{L}\sigma^{\mu\nu}
F_{\mu \nu}b_{R},\\  
{\cal{Q}}_{8}&=&\frac{g_{s} m_{b}}{16 \pi^{2}}\bar{s}_{L}\sigma^{\mu\nu}
G_{\mu\nu}b_{R}.
\end{eqnarray}
Here ${\cal{C}}_{2}(\mu_{0})=1$, and the Wilson coefficients ${\cal{C}}_{7,8}$
can be decomposed according with the particles in the loop,
\begin{eqnarray}
{\cal{C}}_{7}(M_W)={\cal{C}}_{7}^{W}(M_W)+{\cal{C}}_{7}^{H}(M_W)+
{\cal{C}}_{7}^{\chi^{\pm}}(M_W)+ {\cal{C}}_{7}^{\tilde{g} \chi^{0}}(M_W)\\
{\cal{C}}_{8}(M_W)={\cal{C}}_{8}^{W}(M_W)+{\cal{C}}_{8}^{H}(M_W)+
{\cal{C}}_{8}^{\chi^{\pm}}(M_W)+ {\cal{C}}_{8}^{\tilde{g} \chi^{0}}(M_W)
\nonumber
\end{eqnarray} 

Among these contributions, the $W$ penguin is exactly the same as in the SM
and it does not depend on any supersymmetric parameters, it is simply a
function of SM couplings and masses. This contribution is \cite{bertolini}, 
\begin{eqnarray}
{\cal{C}}_{7}^{W}(M_W)&=& -\frac{3}{2} x_{t}\Big( Q_U F_{1}(x_{t})+ 
F_{2}(x_{t})\Big) \nonumber\\
{\cal{C}}_{8}^{W}(M_W)&=&-\frac{3}{2} x_{t} F_{1}(x_{t}) 
\end{eqnarray}
with $x_{t}=m_{t}^{2}/M_{W}^{2}$ and $Q_U$ the charge of the up quarks.
Similarly, in the charged Higgs penguins all the variables are known with the
exception of $M_{h}$. Again this contribution is unchanged by the 
inclusion of the new susy phases,
\begin{eqnarray}
{\cal{C}}_{7}^{H}(M_W)&=&-\frac{x_{t}}{2 x_{h}}
\Big( \cot^{2}\beta (Q_{U} 
F_{1}(x_{t}/x_{h})+ F_{2}(x_{t}/x_{h})) + 
Q_{U} F_{3}(x_{t}/x_{h})+ F_{4}(x_{t}/x_{h}) \Big) \nonumber \\
{\cal{C}}_{8}^{H}(M_W)&=&-\frac{x_{t}}{2 x_{h}}\Big( \cot^{2}\beta  
F_{1}(x_{t}/x_{h})+ F_{3}(x_{t}/x_{h})\Big)  
\end{eqnarray}
where $x_{h}=M_{h}^{2}/M_{W}^{2}$. This contribution gives a sizeable 
correction to the $b \rightarrow s \gamma$ decay that constrains the mass 
of the charged Higgs in two Higgs doublet models or in the MSSM with low 
$\tan \beta$. However, in the case of moderate--large $\tan \beta$,
chargino contributions may partially compensate this charged Higgs 
contribution relaxing the constraints \cite{bsg}.

In addition to the $W^{\pm}$ and charged Higgs contributions analyzed above, 
there are three specifically supersymmetric contributions mediated by gluino, 
neutralino and chargino. In gluino or neutralino diagrams flavor change 
is due to the off--diagonality in the sdown mass matrix.
Being left--right flavor off--diagonal transitions, they are suppressed by 
the mass of the $b$ quark. 
Indeed, smallness of gluino and neutralino contributions has
already been established in \cite{bsg} where it was 
shown that, in the CMSSM, such contributions are roughly one order of
magnitude smaller than the chargino contribution. 

Together with the $W^\pm$ and charged Higgs, the most important supersymmetric 
contribution will be, especially in the large--moderate $\tan \beta$ regime, 
the chargino contribution.  
In the $W$ and charged Higgs contributions, the necessary chirality flip for 
the dipole amplitude is always proportional to $m_b$. However, in the chargino 
penguin the chirality 
flip can be made either through a chargino mass insertion in the loop or 
through an external leg mass insertion proportional to $m_b$. In fact, as 
pointed out in \cite{bertolini}, this enhancement due to $m_{\chi^{i}}/m_{b}$ 
is partially compensated by the presence of the $b$ Yukawa coupling.
Nevertheless, this compensation is only effective for low values of $\tan
\beta$. In terms of the chargino--quark--squark couplings used in the 
previous section, these contributions are,
\begin{eqnarray}
{\cal{C}}_{7}^{\chi^{\pm}}(M_W)=\sum_{k=1}^{6}\sum_{i=1}^{2}
\sum_{\alpha, \beta =u,c,t} \frac{ V_{\alpha b} V_{\beta s}^{*}}
{V_{t b} V_{t s}^{*}} \Big( G^{(\alpha, k) i}{G^{*}}^{(\beta, k) i} 
F_{L}^{7}(z_{k},s_{i}) + \nonumber\\
 \frac{m_{b}}{\sqrt{2} M_W \cos \beta} H^{(\alpha, k) i}
{G^{*}}^{(\beta, k) i} \frac{M_{\chi^{i}}}{m_{b}} F_{R}^{7}(z_{k},s_{i})\Big)
\nonumber \\
{\cal{C}}_{8}^{\chi^{\pm}}(M_W)=\sum_{k=1}^{6}\sum_{i=1}^{2}
\sum_{\alpha, \beta =u,c,t} \frac{ V_{\alpha b} V_{\beta s}^{*}}
{V_{t b} V_{t s}^{*}} \Big( G^{(\alpha, k) i}{G^{*}}^{(\beta, k) i} 
F_{L}^{8}(z_{k},s_{i}) +\nonumber \\
 \frac{m_{b}}{\sqrt{2} M_W \cos \beta} H^{(\alpha, k) i}
{G^{*}}^{(\beta, k) i} \frac{M_{\chi^{i}}}{m_{b}} F_{R}^{8}(z_{k},s_{i})\Big)
\label{charex}
\end{eqnarray} 
with the loop functions defined in Appendix B. Similarly to the situation
for the Wilson Coefficient $C_3$, we can, to a very good approximation,
neglect the presence of intergenerational mixing in the up--squark mass 
matrix \cite{bertolini,cho}, then,
\begin{eqnarray}
{\cal{C}}_{7}^{\chi^{\pm}}(M_W)=\sum_{k=3,6}\sum_{i=1}^{2}
\Big( G^{(3, k) i}{G^{*}}^{(3, k) i} F_{L}^{7}(z_{k},s_{i}) - 
G^{(1, 1) i}{G^{*}}^{(1, 1) i} F_{L}^{7}(z_{1},s_{i})+ \nonumber \\
\frac{m_{\chi^{i}}}{m_{b}} \frac{m_{b}}{\sqrt{2} M_W \cos \beta} (H^{(3, k) i}
{G^{*}}^{(3, k) i} F_{R}^{7}(z_{k},s_{i}) - H^{(1, 1) i}{G^{*}}^{(1, 1) i}
F_{R}^{7}(z_{1},s_{i}) ) \Big) \nonumber \\
{\cal{C}}_{8}^{\chi^{\pm}}(M_W)=\sum_{k=3,6}\sum_{i=1}^{2}
\Big( G^{(3, k) i}{G^{*}}^{(3, k) i} F_{L}^{8}(z_{k},s_{i}) - 
G^{(1, 1) i}{G^{*}}^{(1, 1) i} F_{L}^{8}(z_{1},s_{i}) +\nonumber \\
\frac{m_{\chi^{i}}}{m_{b}} \frac{m_{b}}{\sqrt{2} M_W \cos \beta} (H^{(3, k) i}
{G^{*}}^{(3, k) i} F_{R}^{8}(z_{k},s_{i}) - H^{(1, 1) i}{G^{*}}^{(1, 1) i}
F_{R}^{8}(z_{1},s_{i})) \Big) 
\label{chartot}
\end{eqnarray} 
where, once more, we use CKM unitarity and degeneracy of the first two 
generations of squarks. 

The second term in ${\cal{C}}_{7, 8}$ in Eq. (\ref{char}), which corresponds 
to the chargino mass insertion in the loop, is dominant in the large
$\tan \beta$ regime. Notice that both ${G^{*}}^{(\alpha, k) i}$ and 
$ H^{(\alpha, k) i}$ are products of the squark and chargino mixing matrices
that can be ${\cal O} (1)$ (in the case of flavor--diagonal stop mixings).
Then, for stop and chargino masses around the electroweak scale, this term has 
an extra enhancement of $1/\cos \beta$. This means that, for large 
$\tan \beta$, we can approximate these Wilson Coefficients as,
\begin{eqnarray}
{\cal{C}}_{7}^{\chi^{\pm}}(M_W)=\sum_{k=3,6}\sum_{i=1}^{2}
\frac{m_{\chi^{i}}}{m_{b}} \frac{m_{b}}{\sqrt{2} M_W \cos \beta} (H^{(3, k) i}
{G^{*}}^{(3, k) i} F_{R}^{7}(z_{k},s_{i}) - H^{(1, 1) i}{G^{*}}^{(1, 1) i}
F_{R}^{7}(z_{1},s_{i})) \nonumber \\
{\cal{C}}_{8}^{\chi^{\pm}}(M_W)=\sum_{k=3,6}\sum_{i=1}^{2}
\frac{m_{\chi^{i}}}{m_{b}} \frac{m_{b}}{\sqrt{2} M_W \cos \beta} (H^{(3, k) i}
{G^{*}}^{(3, k) i} F_{R}^{8}(z_{k},s_{i}) - H^{(1, 1) i}{G^{*}}^{(1, 1) i}
F_{R}^{8}(z_{1},s_{i}) )
\label{char}
\end{eqnarray}

\section{$b \rightarrow s \gamma$ and $\varepsilon_{\cal M}$ : Correlated 
analysis}
As we have seen in Section III, chargino contribution to the $C_3$ Wilson 
coefficient, Eq.(\ref{chWCR0}), is the main contribution to indirect CP 
violation of the new supersymmetric phases for large values of
$\tan \beta$. However, if we compare this Wilson coefficient with the 
chargino contribution to the decay $b \rightarrow s \gamma$, 
Eqs. (\ref{chWCR0}) and (\ref{char}), we can see that both chargino 
contributions are deeply related. In fact, if we make a rough approximation 
and assume that the two different loop functions involved are of the 
same order, i.e.,
\begin{eqnarray}
\label{loop-funct}
Y_2(z_k,z_l,s_i,s_j)\approx \sqrt{s_i s_j}\  F_{R}^7(z_k,s_i)\ 
F_{R}^7(z_l,s_j)
\end{eqnarray} 
we would obtain,
\begin{eqnarray}
\label{approx}
C_3(M_W) = ({\cal C}_7(M_W))^2 \frac{m_q^2}{M_W^2}
\end{eqnarray} 
Of course, this cannot be considered as a good approximation. As we can see 
from their explicit expressions in Appendix B, the loop functions are clearly 
different. Anyway, they can be expected to give results of the same 
order of magnitude. So, the order of magnitude of $C_3$ is 
determined by the allowed values of ${\cal C}_7$, as we will explicitly show
below.

To reach this goal, we will follow \cite{kagan-neubert}, where they constrain 
in a model--independent way new physics contributions to the Wilson 
coefficients involved in the $b \rightarrow s \gamma$ decay. In terms of 
these Wilson coefficients, the branching ratio,
$\mbox{BR}(B\rightarrow X_{s} \gamma)$ is,
\begin{eqnarray} 
\label{constrain}
\mbox{BR}(B\rightarrow X_{s} \gamma)\simeq 1.258 + 0.382 |\xi_{7}|^{2} + 
0.015 |\xi_{8}|^{2}+ \\ 1.395 Re[\xi_{7}]+ 0.161 Re[\xi_{8}]
+0.083 Re[\xi_{7}\xi_{8}^{*}] \nonumber
\end{eqnarray}
where $\xi_{a}={\cal{C}}_{a}(M_W)/{\cal{C}}_{a}^{W^{\pm}}(M_W)$. The different
coefficients appearing in Eq.(\ref{constrain}) are the SM renormalization 
group evolved contributions, that must be recovered in the limit $\xi_a=1$.
The numerical values are taken from \cite{kagan-neubert}. We have not taken 
into account the errors associated with the choice of the scale and the 
restrictions on the photon energy that do not modify our conclusions.
Now, using the experimental measure, 
$\mbox{BR}(B\rightarrow X_{s} \gamma)= (3.14 \pm 0.48) \times 10^{-4}$, we can 
constrain the allowed values of the complex variables $\xi_{7}$ and 
$\xi_{8}$. In fact, we can already see from Eq. (\ref{constrain}), that in 
the approximation $\xi_7 \approx \xi_8$ this 
is simply the equation of an ellipse in the $Re[\xi_7]$--$Im[\xi_7]$ 
plane. In the case of supersymmetry with large $\tan \beta$, the new physics 
contribution to $\xi_{7}$ and $\xi_{8}$ will be mainly due to the chargino. 
The allowed values of $\xi_7$ directly constrain then the chargino 
contributions to ${\cal C}_7(M_W)$ and indirectly constrain the values of 
$C_3(M_W)$. 
  
\begin{figure}
\begin{center}
\epsfxsize = 15cm
\epsffile{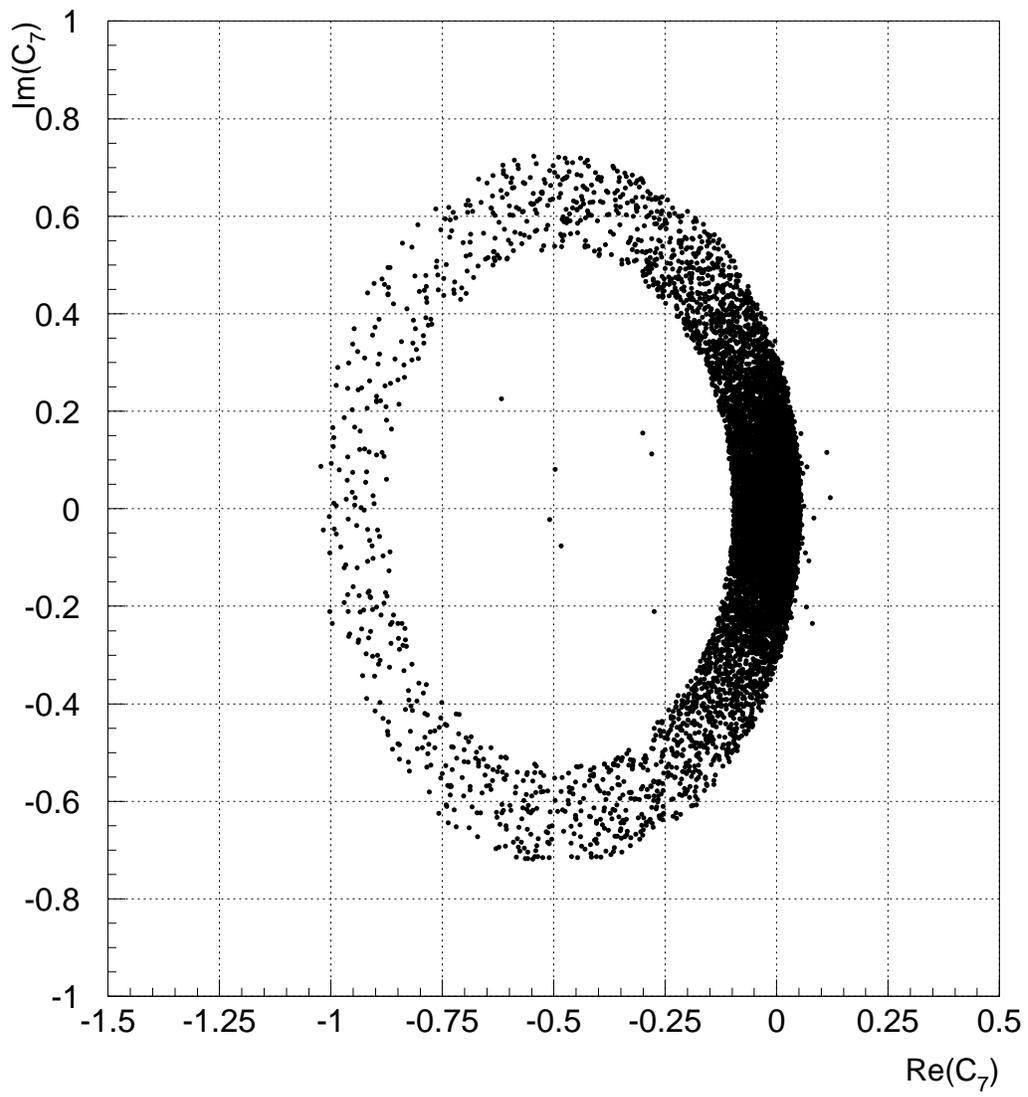}
\leavevmode
\end{center}
\caption{Experimental constraints on the Wilson Coefficient ${\cal C}_7$ }
\label{figc7}
\end{figure}

In figure \ref{figc7}, we show a scatter plot of the allowed values of 
$Re({\cal C}_{7})$ versus $Im({\cal C}_{7})$ in the CMSSM for a fixed value of 
$\tan \beta$ with the constraints from Eq.(\ref{constrain}). Notice, that a 
relatively large value of $\tan \beta$, for example $\tan \beta \gsim 10$, 
is needed to compensate the $W$ and charged Higgs contributions and cover 
the whole allowed area with positive and negative values.
However, the shape of the plot is clearly independent of $\tan \beta$, only
the number of allowed points and its location in the allowed area depend on 
the value considered. In this figure we take $\tan \beta = 40$ 
because only a large value could give rise to observable CP violation 
\cite{fully}. The values of ${\cal C}_7$ and ${\cal C}_8$ used here are the 
values obtained in the CMSSM for a given set of initial conditions.
Although we do not use the approximation $\xi_7 \approx \xi_8$ this does 
not modify the elliptic shape of the plot.

\begin{figure}
\begin{center}
\epsfxsize = 15cm
\epsffile{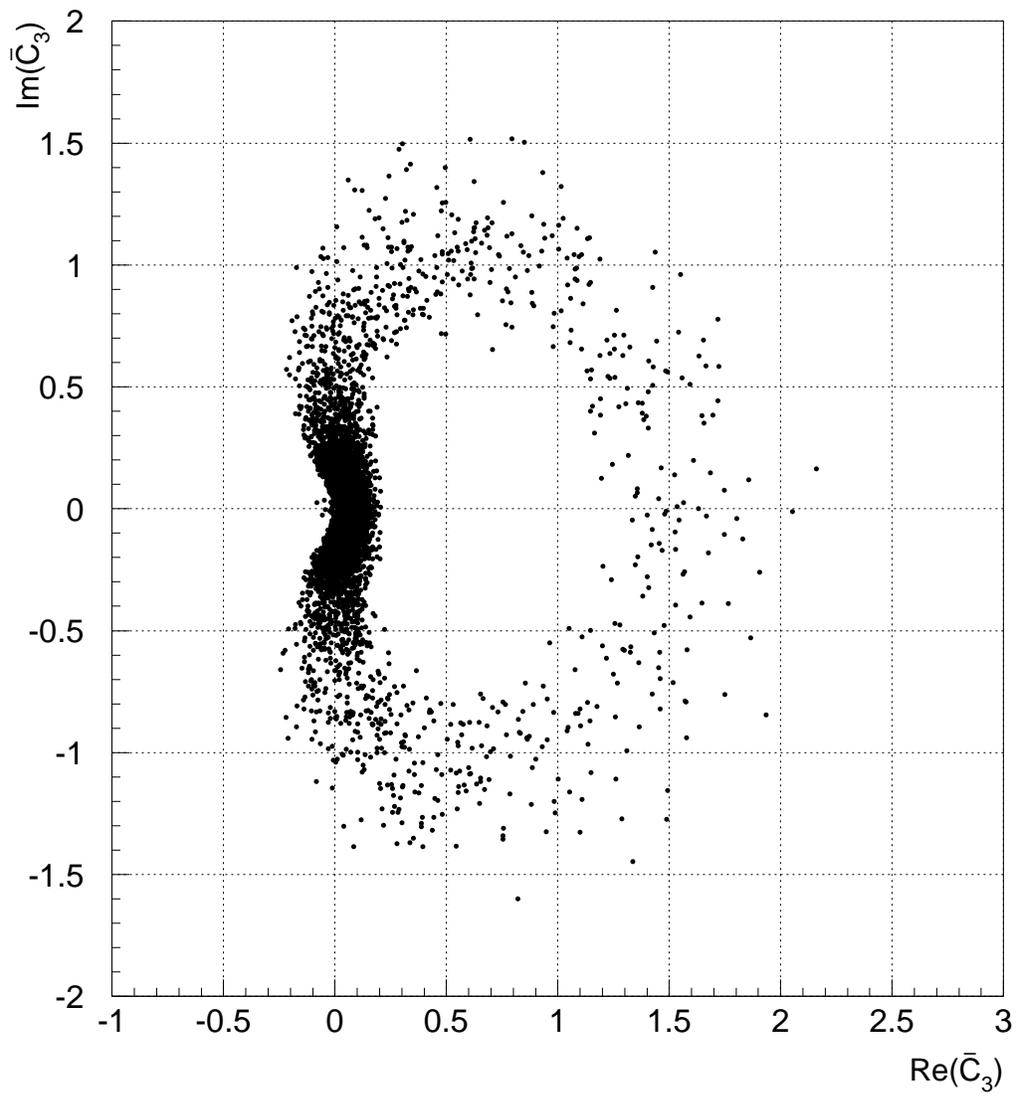}
\leavevmode
\end{center}
\caption{Allowed values for the re--scaled WC $\bar{C}_3$}
\label{figc3}
\end{figure}

Figure \ref{figc3} shows the allowed values for a re--scaled Wilson coefficient
$\bar{C}_3(M_W)= M^2_W/m_q^2 C_3(M_W)$ corresponding to the same allowed 
points of the susy parameter space in figure\ref{figc7} .
As we anticipated previously, the allowed values for $\bar{C}_3$ are close 
to the square of the values of ${\cal C}_7$ in figure \ref{figc7} slightly 
scaled by different values of the loop functions. This is the proof of the 
importance of the $b \rightarrow s \gamma$ constraint on the chargino 
contributions to indirect CP violation.

We can immediately translate this result to a constraint on the size of the 
chargino contributions to $\varepsilon_{\cal M}$. 
\begin{eqnarray}
\label{epscoef}
\varepsilon_{\cal M}=\frac{G_F^2 M_W^2}
{4 \pi^2 \sqrt{2}\ \Delta M_{\cal M}} \frac{(V_{td} V_{tq})^2}{24}
F_{\cal M}^2 M_{\cal M} \eta_3(\mu) B_3(\mu) \frac{M^2_{\cal M}}{m_q^2(\mu)+ 
m_d^2(\mu)} Im[C_3]
\end{eqnarray}
In this expression $M_{{\cal{M}}}$, $\Delta M_{\cal M}$ and $F_{{\cal{M}}}$ 
denote the mass, mass difference and decay constant of the neutral meson 
${\cal{M}}^{0}$. The coefficient $\eta_3(\mu)= 2.93$ \cite{ciuchini} includes 
the RGE effects from $M_W$ to the meson mass scale, $\mu$, and $B_3(\mu)$ is 
the B--parameter associated with the matrix element of the $Q_3$ operator
\cite{ciuchini}.
  
Then, for the $K$ system, using the experimentally measured value 
of $\Delta M_{K}$ we obtain,
\begin{eqnarray}
\label{epsK}
\varepsilon_K^\chi = 1.7 \times 10^{-2} \frac{m_s^2}{M_W^2} Im[\bar{C}_3] 
\approx 0.4 \times 10^{-7} Im[\bar{C}_3]
\end{eqnarray}
Given the allowed values of $\bar{C}_3$ in Fig.\ref{figc3}, this means that
in the CMSSM, even with large susy phases, chargino cannot produce a sizeable 
contribution to $\varepsilon_K$. We have seen in section III that also gluino 
and neutralino give negligible contributions in the CMSSM or in a model 
without off--diagonal soft--breaking terms at the GUT scale. Hence indirect
CP violation in the kaon system will be mainly given by the 
usual SM box and the presence of a CP violating phase in the CKM matrix, 
$\delta_{CKM}$ is still needed. 

The case of $B^0$--$\bar{B}^0$ mixing has a particular interest due to the 
arrival of new data from the B--factories. In fact, as explained at the 
end of section III and in \cite{fully}, in the large $\tan \beta$ regime 
chargino contributions to indirect CP violation can be very important.
However, for any value of $\tan \beta$ we must satisfy the bounds from the
$b \rightarrow s \gamma$ decay. So, if we apply these constraints to the 
$B^0$--$\bar{B}^0$ mixing,
\begin{eqnarray}
\label{epsB}
\varepsilon_B^\chi = 0.17 \frac{m_b^2}{M_W^2} Im[\bar{C}_3] 
\approx 0.5 \times 10^{-3} Im[\bar{C}_3]
\end{eqnarray}
where once again, with the allowed values of Fig.\ref{figc3} we get a very 
small contribution to CP violation in the mixing.
We must take into account that the mixing--induced CP phase, $\theta_M$, 
measurable in $B^0$ CP asymmetries, is related to $\varepsilon_B$ by 
$\theta_M=\arcsin\{2 \sqrt{2} \cdot \varepsilon_B \}$. The expected 
sensitivities on the CP phases at the B factories are around $\pm 0.1$ 
radians, so this supersymmetric chargino contribution will be completely 
out of reach. Gluino and neutralino contributions to indirect CP violation
can also be discarded in the CMSSM. Once again we have to conclude that no
new contributions to indirect CP violation from the new susy phases will be 
observable in $B^0$ CP asymmetries in the framework of the CMSSM. 
Recently, CDF \cite{CDF} has provided preliminary indications that 
$\sin 2 \beta$ is in agreement with the SM predictions. Clearly, from the 
above result, Eq. (\ref{epsB}), it appears that the CMSSM contribution is 
too small by itself to account for this result.

\section{Conclusions}

In this work, the effects of non-vanishing supersymmetric phases on indirect CP
violation in $K$ and $B$ systems have been analyzed within the CMSSM.
We have found that operators involving only left--handed external quarks are 
not sensitive to these new phases at an observable level. This is due to the 
absence of intergenerational mixings beyond those originated from the CKM 
matrix. 
On the contrary, operators involving both right and left--handed quarks are 
in general complex, even in the absence of $\delta_{CKM}$, and could be 
relevant in the large $\tan \beta$ regime. However, we have shown that these 
contributions are deeply related with the 
$\mbox{BR}(B\rightarrow X_{s} \gamma)$ decay. So, taking into account the 
constraints coming from this decay these contributions also turn out to be  
too small to be measured experimentally.

Although these conclusions are specific for indirect CP violation, they could
also be implemented for chargino mediated direct CP violation in the decays.
Again, in these decays the same chargino--quark--squark couplings are involved
and we can also expect a big impact of the $b \rightarrow s \gamma$ constrain. 
In fact, the conclusions reached in this paper are far more general.
The correlation between $b \rightarrow s \gamma$ and susy induced indirect CP
violation exists in any supersymmetric model with sufficiently small 
intergenerational mixings in the sfermion mass matrices.
This would include specifically all the models without new flavor structures
beyond the usual CKM matrix at the GUT scale and simplified models as the one 
the authors used in
\cite{fully}.

In summary, concerning the simpler supersymmetric models, like CMSSM, the 
constraints coming from $\mbox{BR}(B\rightarrow X_{s} \gamma)$ decay are 
sufficient to rule out pure supersymmetric indirect CP violation in $K$ and 
$B$ systems, even in the absence of any electric dipole moment constraints. 
This has very important consequences for the supergravity induced models
where a cancellation between different supersymmetric contributions allows 
large supersymmetric phases while respecting EDM bounds \cite{cancel}.
In these models, even in the regions of parameter space where this cancellation
occurs, no observable effect of the large susy phases will appear on indirect
CP violation experiments.
However, as pointed out by Baek and Ko \cite{korea}, these phases would still 
be observable in CP asymmetries in the $b \rightarrow s \gamma$ decay.
  
All this means that the presence of large susy phases is not sufficient to 
produce observable effects at the low energy experiments. In particular,
new sources of flavor change beyond the usual CKM matrix are needed.
And so, any deviation from the SM expectations at indirect CP violation
experiments due to supersymmetry should be taken as a sign of 
non--universality of the soft breaking terms. In this context one recalls
the recent studies on superstring compactifications with non--universal 
gaugino masses \cite{kane}.

\acknowledgments{We thank S. Bertolini for useful discussions and S. Baek 
and P. Ko for fruitful mail exchange. D.A.D. thanks  P. Langacker for
his helpful comments concerning the integration of RGE's.
The work of A.M. was partially supported by the European TMR Project
``Beyond the Standard Model'' contract N. ERBFMRX CT96 0090; O.V. 
acknowledges financial support from a Marie Curie EC grant 
(TMR-ERBFMBI CT98 3087).} 
\newpage
\section*{Appendix A. Integration of RGE's in CMSSM}
\setcounter{equation}{0}
\def\theequation{A.\arabic{equation}}

In this Appendix we describe the new features of the integration of RGE's in 
the CMSSM with non-vanishing susy phases relevant to our analysis. 
The complete matrix form of the RG equations can be found in \cite{bertolini}. 
Using their notation and conventions, with the only change of $A_q= m Y_q^A$,
we will mainly concentrate on the left--left scalar--quark mass matrix and 
the tri--linear soft breaking coupling evolution,
\begin{eqnarray}
\label{mqrge}
\frac{d m_Q^2}{d t} =& \left(\frac{16}{3}\tilde{\alpha}_3 M^2_3 + 3 
\tilde{\alpha}_2 M^2_2 + \frac{1}{9}\tilde{\alpha}_1 M^2_1\right){\bf 1} -
\nonumber\\
& \frac{1}{2}\left[ \tilde{Y}_U\tilde{Y}_U^\dagger m_Q^2 + m_Q^2 \tilde{Y}_U
\tilde{Y}_U^\dagger + 2 ( \tilde{Y}_U m_U^2 \tilde{Y}_U^\dagger +
\bar{\mu}^2_2  \tilde{Y}_U\tilde{Y}_U^\dagger +  \tilde{A}_U
\tilde{A}_U^\dagger) \right] \nonumber +\\
& \frac{1}{2}[ \tilde{Y}_D\tilde{Y}_D^\dagger m_Q^2 + m_Q^2 \tilde{Y}_D
\tilde{Y}_D^\dagger + 2 ( \tilde{Y}_D m_D^2 \tilde{Y}_D^\dagger +
\bar{\mu}^2_1  \tilde{Y}_D\tilde{Y}_D^\dagger +  \tilde{A}_D
\tilde{A}_D^\dagger) ] 
\end{eqnarray}

\begin{eqnarray}
\label{Aurge}
\frac{d \tilde{A}_U}{d t} =& \frac{1}{2} \left(\frac{16}{3}\tilde{\alpha}_3 + 
3 \tilde{\alpha}_2 + \frac{1}{9}\tilde{\alpha}_1 \right) \tilde{A}_U -
\left(\frac{16}{3}\tilde{\alpha}_3 M_3 + 3 \tilde{\alpha}_2 M_2 + 
\frac{1}{9}\tilde{\alpha}_1 M_1\right) \tilde{Y}_U - \nonumber \\
& \frac{1}{2} [ 4\tilde{A}_U \tilde{Y}_U^\dagger\tilde{Y}_U + 
6 Tr(\tilde{A}_U \tilde{Y}_U^\dagger)\tilde{Y}_U + 5 \tilde{Y}_U 
\tilde{Y}_U^\dagger \tilde{A}_U + 3 Tr(\tilde{Y}_U \tilde{Y}_U^\dagger)
\tilde{A}_U +\nonumber \\
&2 \tilde{A}_D \tilde{Y}_D^\dagger\tilde{Y}_U + 
\tilde{Y}_D \tilde{Y}_D^\dagger \tilde{A}_U ] 
\end{eqnarray}

\begin{eqnarray}
\label{Adrge}
\frac{d \tilde{A}_D}{d t} =& \frac{1}{2} \left(\frac{16}{3}\tilde{\alpha}_3 + 
3 \tilde{\alpha}_2 + \frac{1}{9}\tilde{\alpha}_1 \right) \tilde{A}_D -
\left(\frac{16}{3}\tilde{\alpha}_3 M_3 + 3 \tilde{\alpha}_2 M_2 + 
\frac{1}{9}\tilde{\alpha}_1 M_1\right) \tilde{Y}_D - \nonumber \\
& \frac{1}{2} [ 4\tilde{A}_D \tilde{Y}_D^\dagger\tilde{Y}_D + 
6 Tr(\tilde{A}_D \tilde{Y}_D^\dagger)\tilde{Y}_D + 5 \tilde{Y}_D 
\tilde{Y}_D^\dagger \tilde{A}_D + 3 Tr(\tilde{Y}_D \tilde{Y}_D^\dagger)
\tilde{A}_D +\nonumber \\
&2 \tilde{A}_U \tilde{Y}_U^\dagger\tilde{Y}_D + 
\tilde{Y}_U \tilde{Y}_U^\dagger \tilde{A}_D + 2 Tr(\tilde{A}_E 
\tilde{Y}_E^\dagger)\tilde{Y}_D + Tr(\tilde{Y}_E \tilde{Y}_E^\dagger)
\tilde{A}_D ] 
\end{eqnarray}
 
Except for the Yukawa coupling matrices, the RGE's of all other quantities 
are linear \cite{bertolini}. This means, in particular, that RGE's of all soft 
masses, though coupled, can be solved as a linear combination of the
GUT--scale parameters $m_{0}$, $A_{0}e^{i\phi_{A}}$ and $M_{1/2}$ at any 
scale below $M_{G}$. However, one notices that the initial conditions on the 
trilinear couplings require the knowledge of the particular Yukawa texture at 
the unification scale. To do this, we numerically integrate the Yukawa RGE 
with a given value of $\tan \beta$ and in terms of the fermion masses and the 
CKM matrix. Specifying the GUT--scale initial conditions in this way, it is 
straightforward to compute all soft masses at $M_{W}$ for arbitrary values of 
$m_{0}$, $A_{0}e^{i\phi_{A}}$ and $M_{1/2}$. 
Thanks to the linearity of the corresponding RGE's the soft masses at $M_{W}$ 
admit the following expansion,
\begin{eqnarray}
\label{etas}
A_{U,D}(M_{Z})&=&\alpha^{A}_{U,D}\,A_{0} e^{i\phi_{A}}+\alpha^{g}_{U,D} 
M_{1/2}\nonumber\\
m_{Q,U,D}^{2}(M_{Z})&=&\eta^{m}_{Q,U,D}\,m_{0}^{2}+\eta^{A}_{Q,U,D}\,A_{0}^{2}
+\eta^{g}_{Q,U,D}\,M_{1/2}^{2}\\&+&
\Big(\eta^{(g\,A)}_{Q,U,D}\,  e^{i\phi_{A}}+ \eta^{(g\,A)\, T}_{Q,U,D}\,  
e^{-i\phi_{A}}\Big) A_{0} M_{1/2} \nonumber
\end{eqnarray}
where the coefficients $\alpha$ and $\eta$ are $3\times 3$ matrices with 
real numerical entries. One notices that the matrices $m_{Q,U,D}^{2}(M_{Z})$ 
would be completely real were it not for the non--symmetric terms in the matrix
$\eta^{g\,A}_{Q,U,D}$. However, it will be seen from the specific examples 
that this matrix remains nearly symmetric, and thus, CP--violating entries 
$m_{Q,U,D}^{2}(M_{Z})$ are extremely suppressed. Moreover, one notices that
$A_{U,D}(M_{Z})$ carries, in general, large CP violating phases; however, 
these terms are effective only for intragenerational $LR$-type mixings. Hence, 
this particular observation shows the importance of chargino contributions 
for CP violation in FCNC processes, as explained in sec. II. 

As mentioned before, due to the non--linearity of the RGE's for Yukawa 
matrices, it is not possible to give a fully analytic solution for the soft 
mass parameters. Nevertheless, once we fix $\tan\beta$, we can numerically
integrate the Yukawa RGE. Therefore, below we give semi--analytic solutions 
of RGE's for $\tan\beta=2$ and $\tan\beta=40$ to illustrate the small and 
large $\tan\beta$ regimes. 

Fixing $\tan \beta=2$, we get for the relevant $\eta$ matrices in 
Eq.(\ref{etas}),

\begin{eqnarray}
\eta^{g}_{Q}&=&\left (\begin{array}{c c c}
7.07& 2.79\times 10^{-4}& -7.02\times 10^{-3}\\
2.79\times 10^{-4}& 7.07& 4.92\times 10^{-2}\\
-7.02\times 10^{-3}& 4.92\times 10^{-2}& 5.74\end{array}\right)\\
\frac{1}{2}( \eta^{(g\,A)}_{Q} + \eta^{(g\,A)\, T}_{Q}) &=&\left(
\begin{array}{c c c}
5.34\times 10^{-6}& -3.44\times 10^{-5}& 7.90\times 10^{-4}\\
-3.44\times 10^{-5}& 2.29\times 10^{-4}& -5.52\times 10^{-3}\\
7.90\times 10^{-4}& -5.52\times 10^{-3}& 0.15\end{array}\right)\\
\frac{1}{2}(\eta^{(g\,A)}_{Q} - \eta^{(g\,A)\, T}_{Q}) &=&
\left(\begin{array}{c c c}
0 & 0 & 1.34\times 10^{-8}\\
0 & 0 & -8.55\times 10^{-8}\\
-1.34 \times 10^{-8}& 8.55\times 10^{-8}& 0 \end{array}\right)
\end{eqnarray}
where the vanishing off--diagonal entries in the last matrix mean values 
smaller than $10^{-10}$ in absolute magnitude.
Among the matrices involved in Eq.(\ref{etas}), $\eta^g$ is always the largest
one for similar values of $M_{1/2}$ and $m_0$. So, it sets the scale of the 
matrix element while $\eta^{(g A)}$ is the only one that can produce an
imaginary part. Hence, we do not specify the other $\eta$--matrices, 
which are not important for our discussion.

Once we obtain the $m_Q (M_W)$ matrix with the help of Eq.(\ref{etas}) we
can get the values of the ${M^{(u)}_{LL}}^2$ and ${M^{(d)}_{LL}}^2$ in the
SCKM basis, that give the size of flavor change in the squark mass matrices 
compared with the diagonal elements. For $\tan\beta=2$, those elements of
the squark mass--squared matrix causing $LL$ transitions between first 
and second,
as well as second and third generations, are given by 

\begin{eqnarray}
\label{off2}
{({M_{LL}^{(u)}}^{2})}_{1 2}&=&-2.79\times 10^{-7}\, m_{0}^{2}- 
9.30\times 10^{-8}\, A_{0}^{2} -1.17\times 10^{-6}\,
M_{1/2}^{2}\nonumber\\ &+& 8.15\times 10^{-7}\, A_{0} M_{1/2} \cos\phi_{A}\\
{({M_{LL}^{(u)}}^{2})}_{2 3}&=&-4.07\times 10^{-5}\, m_{0}^{2} 
-1.15\times 10^{-5}\, A_{0}^{2} - 1.61\times
10^{-4}\, M_{1/2}^{2}\nonumber\\&+&1\times 10^{-4}\, A_{0} M_{1/2} 
\cos\phi_{A}
-\,1.71\times 10^{-7}\, A_{0} M_{1/2}\ i \sin\phi_{A}\\
{({M_{LL}^{(d)}}^{2})}_{1 2}&=&9.38\times 10^{-5}\, m_{0}^{2}+
3.75\times 10^{-6}\, A_{0}^{2} + 2.79\times 10^{-4}\, M_{1/2}^{2}\nonumber\\
&+& 6.87\times 10^{-5}\, A_{0} M_{1/2} \cos\phi_{A}\\
{({M_{LL}^{(d)}}^{2})}_{2 3}&=&1.67\times 10^{-2}\, m_{0}^{2}+
5.32\times 10^{-4}\, A_{0}^{2} + 4.91\times
10^{-2}\, M_{1/2}^{2}\nonumber\\&-&1.1\times 10^{-2}\, 
A_{0} M_{1/2} \cos\phi_{A}
- 1.70\times 10^{-7}\, A_{0} M_{1/2}\ i \sin\phi_{A}\,.
\end{eqnarray}

Now, we repeat the same quantities above for $\tan\beta=40$:
 
\begin{eqnarray}
\eta^{g}_{Q}&=&\left (\begin{array}{c c c}
7.07& 2.44\times 10^{-4}& -5.80\times 10^{-3}\\
2.44\times 10^{-4}& 7.07& 4.06\times 10^{-2}\\
-5.80\times 10^{-3}& 4.06\times 10^{-2}& 4.97\end{array}\right)\\
\frac{1}{2}( \eta^{(g\,A)}_{Q} + \eta^{(g\,A)\, T}_{Q}) &=&\left(
\begin{array}{c c c}
8.32\times 10^{-6}& -4.57\times 10^{-5}& 7.82\times 10^{-4}\\
-4.57\times 10^{-5}& 5.20\times 10^{-4}& -5.47\times 10^{-3}\\
7.82\times 10^{-4}& -5.47\times 10^{-3}& 0.22\end{array}\right)\\
\frac{1}{2}(\eta^{(g\,A)}_{Q} - \eta^{(g\,A)\, T}_{Q}) &=&
\left(\begin{array}{c c c}
0 & 0 & -1.64\times 10^{-6}\\
0 & 0 &  1.14\times 10^{-5}\\
1.64 \times 10^{-6}& -1.14\times 10^{-5}& 0 \end{array}\right)\,.
\end{eqnarray}

\begin{eqnarray}
\label{off40}
{({M_{LL}^{2}}^{(u)})}_{1 2}&=&-8.77\times 10^{-5}\, m_{0}^{2}- 2.77
\times 10^{-5}\, A_{0}^{2} -3.0\times 10^{-4}\,
M_{1/2}^{2}\nonumber\\ &+& 1.21\times 10^{-4}\, A_{0} M_{1/2} \cos\phi_{A}+i
\,1.1\times 10^{-10}\, A_{0} M_{1/2} \sin\phi_{A}\\
{({M_{LL}^{2}}^{(u)})}_{2 3}&=&-1.28\times 10^{-2}\, m_{0}^{2}-2.70\times 
10^{-3}\, A_{0}^{2} -3.77\times
10^{-2}\, M_{1/2}^{2}\nonumber\\&+&5.67\times 10^{-3}\, A_{0} M_{1/2} 
\cos\phi_{A}
+i\,2.30\times 10^{-5}\, A_{0} M_{1/2} \sin\phi_{A}\\
{({M_{LL}^{2}}^{(d)})}_{1 2}&=&7.51\times 10^{-5}\, m_{0}^{2}+7.74
\times 10^{-6}\, A_{0}^{2} + 2.44\times 10^{-4}\, M_{1/2}^{2}\nonumber\\
&-& 9.13\times 10^{-5}\, A_{0} M_{1/2} \cos\phi_{A}\\
{({M_{LL}^{2}}^{(d)})}_{2 3}&=&1.34\times 10^{-2}\, m_{0}^{2}+7.84
\times 10^{-4}\, A_{0}^{2} + 4.05\times
10^{-2}\, M_{1/2}^{2}\nonumber\\&-&1.1\times 10^{-2}\, A_{0} M_{1/2} 
\cos\phi_{A}
+i\,2.28\times 10^{-5}\, A_{0} M_{1/2} \sin\phi_{A}\,.
\end{eqnarray}

A comparison of the corresponding quantities in $\tan\beta=2$ and 
$\tan\beta=40$ cases reveals the sensitivity of the results on $\tan\beta$. 
As explained in Sec. II, $Y_{U}(M_{Z})$ remains nearly unchanged while 
$Y_{D}(M_{Z})$ assumes an order of magnitude enhancement as $\tan\beta$ varies
from 2 to 40. This change in $Y_{D}(M_{Z})$ affects various quantities as
dictated by the differential equations (A.1 -- A.3).
\newpage
\section*{Appendix B. Loop functions}
\setcounter{equation}{0}
\def\theequation{B.\arabic{equation}}
In this appendix we collect the different loop function used in the text.
The functions $Y_1$ and $Y_2$ entering $B$--$\bar{B}$ and $K$--$\bar{K}$
mixings are given by 
\begin{eqnarray}
\lefteqn{ Y_1(a,b,c,d) = }              \nonumber  \\
 & &  {a^2\over (b-a)(c-a)(d-a)}\ln{a}
     +{b^2\over (a-b)(c-b)(d-b)}\ln{b} 
                                             \nonumber \\  
 & & +{c^2\over (a-c)(b-c)(d-c)}\ln{c}
     +{d^2\over (a-d)(b-d)(c-d)}\ln{d}
\end{eqnarray}
and
\begin{eqnarray}
\lefteqn{ Y_2(a,b,c,d) = }               \nonumber \\
 & & \sqrt{4 c d}\Bigg[{a\over (b-a)(c-a)(d-a)}\ln {a}
     +{b\over (a-b)(c-b)(d-b)}\ln{b}
                                              \nonumber \\
 & & +{c\over (a-c)(b-c)(d-c)}\ln{c}
     +{d\over (a-d)(b-d)(c-d)}\ln{d}\Bigg]\,.
\end{eqnarray}

For the analysis of $b\rightarrow s \gamma$ branching ratio the following 
loop functions are relevant:

\begin{eqnarray}
  F_1(x)   &=&
        \frac{1}{12(x-1)^4}(x^3-6x^2+3x+2+6x\ln x),
\\
  F_2(x)   &=&
        \frac{1}{12(x-1)^4}(2x^3+3x^2-6x+1-6x^2\ln x),
\\
  F_3(x)   &=&
        \frac{1}{2(x-1)^3}(x^2-4x+3+2\ln x),
\\
  F_4(x)   &=&
        \frac{1}{2(x-1)^3}(x^2-1-2x\ln x),
\\
  F_{L}^{7}(x,y) &=&
        \frac{1}{x}
        \left[Q_{U}F_{2}(y/x)+F_{1}(y/x)\right],
\\
  F_{R}^{7}(x,y) &=&
        \frac{1}{x}  
        \left[Q_{U}F_{4}(y/x)+F_{3}(y/x)\right],                      
\\
  F_{L}^{8}(x,y) &=&
        \frac{1}{x}F_{2}(y/x),                      
\\
  F_{R}^{8}(x,y) &=&
        \frac{1}{x} F_{4}(y/x)\,.
\end{eqnarray}

\clearpage
\end{document}